# On the potential of Hg-Photo-CVD process for the low temperature growth of nano-crystalline silicon
## (Topical review)

## A. BARHDADI

*Laboratoire de Physique des Semi-conducteurs et de l'Energie Solaire (P.S.E.S.)*
*Ecole Normale Supérieure, BP. 5118, Takaddoum, Rabat-10000, Morocco*
Phone: (212) 37 75 12 29  or  (212) 37 75 22 61 or  (212) 64 93 68 15    Fax: (212) 37 75 00 47
E-mails: barhdadi@ictp.trieste.it   or   abdelbar@fsr.ac.ma

&

*Unité de Formation et de Recherche en Physique de la Matière Condensée et Modélisation Statistique des Systèmes*
*Faculté des Sciences, Université Mohammed V-Agdal, B.P. 1014, Rabat, Morocco*

&

*The Abdus Salam International Centre for Theoretical Physics (ICTP)*
*Strada Costiera 11, 34014, Trieste, Italy*

## Abstract

Mercury-Sensitized Photo-Assisted Chemical Vapor Deposition (Hg-Photo-CVD) technique opens new possibilities for reducing thin film growth temperature and producing novel semiconductor materials suitable for the future generation of high efficiency thin film solar cells onto low cost flexible plastic substrates. This paper provides an overview of this technique, with the emphasis on its potential in low temperature elaboration of nano-crystalline silicon for the development of thin films photovoltaic technology.

## Résumé

La technique de dépôt photochimique en phase vapeur sensibilisé au mercure (Hg-Photo-CVD) ouvre de nouvelles possibilités pour réduire la température de croissance des couches minces et produire de nouveaux matériaux semiconducteurs convenables pour la future génération de photopiles, de haut rendement et petit prix, réalisées en couches minces déposées sur des substrats flexibles en plastique. Cet article fournit une vue d'ensemble sur cette technique, avec l'emphase sur son potentiel dans l'élaboration du silicium nanocristallin à faible température pour le développement de la technologie photovoltaïque sur les couches minces.

# I- INTRODUCTION

The key driver for the future generation photovoltaic is the cost reduction to about $ 1/Wp at the system level. In order to achieve this ambitious goal, the focus of photovoltaic R&D is currently in developing novel thin film technologies, involving new semiconductor materials and solar cell devices, as well as pursuing radically new concepts that can significantly improve the conversion efficiency and, at the same time, reduce notably the photovoltaic price [1, 2]. In this perspective, nano-crystalline silicon (nc-Si) has recently raised a great deal of interest and emerged as a potential new semiconductor candidate for the development of low cost efficient solar cells [3, 4]. Because of its quite particular structure and most interesting properties, nc-Si is proposed to be commonly used in developing new potential applications not only in photovoltaic but also in optoelectronic and microelectronic technologies (flat panel displays and imaging devices, thin-film transistors, large area sensors, etc.) [5, 6].

Nc-Si, with a typical grain size of 10 nm, was first reported in 1968 [7]. For many years ago, because of the famous correlation between solar cell efficiency and crystalline silicon grain size [8], no one believed in the possibility of using this material for any photovoltaic application. However, in spite of its nano-scaled grains, recent experimental research developments reveal the possibility of obtaining high conversion efficiencies from solar cells based on nc-Si [9-13] thanks to the low temperature thin film deposition process under heavily hydrogen atmosphere [3, 4]. The predominant explanation of this arises mainly from the well-known hydrogen passivation of electrically active defects as well as harmful contaminants [14-22].

As in the case of amorphous silicon (a-Si), but under slightly modified conditions, nc-Si material is commonly deposited in the form of thin films at low temperature usually in the range of 200-400°C [3]. Most often, the deposition process results a plasma of silane ($SiH_4$), or other appropriate silicon carrier gases, heavily diluted with hydrogen ($H_2$); so that there is a high concentration of atomic hydrogen in the plasma. Of course, the low growth temperature process reduces significantly the grain size of the deposited film, nevertheless a large amount of hydrogen atoms remains inside the film and then passivates the electrical activity of dangling bond related defects as in the case of hydrogenated amorphous silicon (a-Si:H) [23]. The exact means by which nc-Si, rather than a-Si, forms has been the subject of long scientific debate. The most popular models proposed to explain the nc-Si deposition process are called the surface diffusion [24, 25] and the selective etching [26-28]. In the first one, high hydrogen atoms flux impinging on the surface is thought to enhance the mobility of the deposition precursors. In the selective etching model, both nc-Si and a-Si solid phases are assumed to be deposited simultaneously, but the hydrogen atoms as well as hydrogen-based radicals of the plasma etches away a-Si phase more quickly, leaving behind mostly nc-Si [28]. Once formed, the two solid phases can then interconvert via the gas phase. Even if the two models are plausible, their mechanisms do not explain how thin a-Si:H films crystallize at temperatures much lower than those required for thermal crystallization, upon post-deposition exposure to hydrogen atoms created through a plasma dissociation of $H_2$ [26-32]. An alternative model, called "chemical annealing", has been then proposed as a believable mechanism explaining the process. In this new model, atomic hydrogen permeates the a-Si:H film and, through its insertion into strained Si-Si bonds, annihilates the strains and lowers the energy barrier to the rearrangement of the silicon network into a more stable, more ordered nano-crystalline lattice [32-35]. Nevertheless, until now, the precise atomic-scale mechanism for this process has not yet fully elucidated even if the simulation work performed by Sriraman et al. provides an important contribution in this way [36].

The performances of nc-Si solar cells are intimately related to the film structure determined by the preparation method and the experimental conditions adopted during the deposition process. Various deposition techniques have been employed [37-42]. Among them, Plasma-Enhanced Chemical Vapor Deposition (PECVD) [42-47] still the most commonly used for the direct growth of nc-Si thin films on low substrate temperature, and $SiH_4$ heavily diluted with $H_2$ are generally used as the reactants. However, PECVD has yet two serious drawbacks: the surface damages which result from the impinging charged particles with high energy of the plasma, and the impurities incorporation which results from sputtering because of the high potential difference between substrate and electrodes in the reactor [48]. Recently, Myong et al. [49] reported that Mercury-Sensitized Photo-Assisted Chemical Vapor Deposition (Hg-Photo-CVD) is not only an efficient method for deposing high quality nc-Si thin films at temperature as lower as 120°C, but also a promising technique for developing the future generation of thin film solar cells fabricated onto low cost flexible plastic substrates. At home laboratory, my group and I have been very charmed by this result and we have immediately agreed to develop this new topic, among our current research activities, to eventually perform some significant original contributions. So, presently, we almost finished the development of a new thin film deposition set-up for the technique of Hg-Photo-CVD [50-52]. This set up is quite similar to that performed by our collaborator Aka some years ago in France [53-56]. The specific research program that we are planning to carry out on this new system registers within the framework of a research project financed by the Moroccan ministry of high education and scientific research. It mainly consists in depositing good quality a-Si:H and nc-Si thin films in the perspective to contribute in the development of the future generation of low cost powerful terrestrial solar cells. To conduct well this new research program, we initially performed an extensive bibliographical work aiming to seek and join together most of scientific information and technical data published by the specialists of the subject. After examining and analyzing these literature data, we noted some fundamental points characterizing particularly nc-Si material and Hg-Photo-CVD technique. That is what we are trying to recall and review through the bibliographical synthesis we are proposing in this paper. Our motivation is to develop a new synthesis work dedicated mainly to outline the great photovoltaic potential of the low temperature deposition process of nc-Si thin films for the development of the new generation of high efficiency low cost solar cells.

## II- Background information on nc-Si

Currently, non-single-crystalline silicon is referred to by a variety of names depending on its single-crystal silicon particle (most frequently called crystallite or grain) size. So, we may distinguish between nano-crystalline (nc-Si), micro-crystalline (μc-Si), poly-crystalline (pc-Si) and multi-crystalline (mc-Si) silicon, if ordered by the grain size. In fact, these are roughly discriminated by the crystalline size in the material, so most often nc-Si implies crystallites having a typical average size of 10 nm (figure 1) with only a small dispersion [57]. Commonly, these crystallites are either abutted against each other [58] or embedded in an amorphous silicon matrix and present an inhomogeneous distribution [57, 59]. This nano-structured silicon material is also called polymorphous silicon when various crystalline phases coexist in the nano-crystal [59]. Nc-Si material is then similar to a-Si because has an amorphous solid phase, but different because has at the same time a distribution of nano-grains of crystalline silicon (c-Si) within the amorphous phase. This is in contrast to pc-Si or mc-Si which consists solely of crystalline silicon grains, separated by grain boundaries.

Low temperature synthesis of nc-Si was first reported by Vaprec et al. in 1968 [7]. A crystalline silicon source at room temperature is etched out by atomic hydrogen generated by remote plasma and the etching products are transferred to a heated substrate on which a deposition of a silicon thin film occurs [61]. The chemical equilibrium between the crystalline silicon source and a deposited film on the substrate is established to form nc-Si. The chemical transport experiment is similar in the principle to the currently employed plasma CVD process (figure 2) in the respect that deposition precursors are $SiH_4$ related radicals such as $SiH_x$ and that atomic hydrogen plays a crucial role, because these species are also generated by the plasma from a gas mixture of $SiH_4$ and $H_2$.

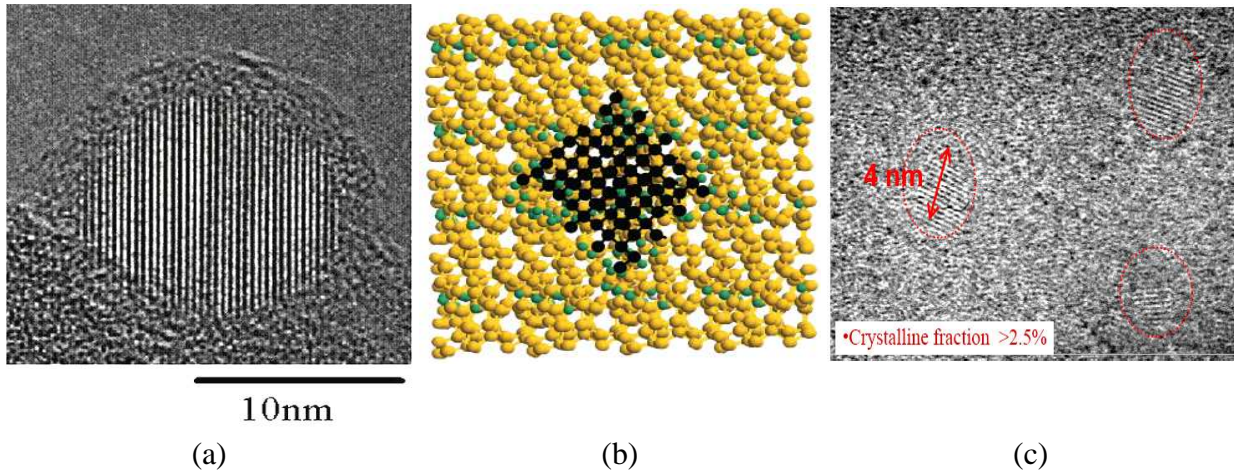

(a) (b) (c)

*Figure 1: Nano-crystalline silicon particle*
*a) Perfect single-crystal silicon particle with a typically diameter of about 10 nm covered embedded in amorphous silicon material (after [57]).*
*b) A picture of nano-crystalline silicon structure from a molecular dynamics simulation with nanometer size crystallite embedded in an amorphous matrix. The silicon atoms in the crystallite are black and are light-colored for the amorphous matrix. The hydrogen atoms are small lighter spheres (after [60]).*
*c) nc-Si cystallites embedded in an amorphous silicon matrix and present an inhomogeneous distribution (after [59]).*

The deposition of nc-Si using plasma CVD of $SiH_4$ started in 1979-1980 [62]. In 1979, an anomalously high conductivity of phosphorous doped a-Si:H thin films deposited from $SiF_4$ source gas has been reported [62]. This high conductivity has been first explained by a more ideal silicon network formed by the use of $SiF_4$ gas rather than $SiH_4$ [62]. In 1980, after a fine examination of the films, the presence of very small silicon crystallites has been revealed [63]. At approximately the same time, the discovery of μc-Si synthesized by plasma CVD using high $H_2$ dilution of $SiH_4$ has been reported by several Japanese groups [38-40]. Usui et al. [64] reported the heavy phosphorous doped μc-Si thin films and Hamasaki et al. [65] as well as Matsuda et al. [66] separately reported the microcrystalline solid phase formation by deposition at low temperature. In 1990, silicon-based visible light emission devices were stimulated by electrochemically prepared porous silicon [67], and this study was followed by many others concerning nc-Si [68]. The use of nc-Si in solar cells application was attempted for the photocell transmitting window layer due to its high conductivity and its low absorption coefficient in the visible region [69, 70]. Thin film solar cells need indeed window layers with high electrical conductivity along with a wide optical band gap. Both requirements can be met by implementing inhomogeneous silicon-carbon thin films containing nano-crystalline and amorphous structure phases. The silicon-carbon amorphous phase provides a wide gap optical transmittance, while the nano-crystalline phase fraction provides a high electrical conductivity [71]. However, nc-Si solar cell was not successful until the Neuchatel group reported 7% efficiency using gas purification and the very high frequency plasma technique [9, 72]. At present, an efficiency of more that 10% is obtained for single junction solar cell [11], and the initial efficiency of 14.1% for double junction with an a-Si:H top cell [12].

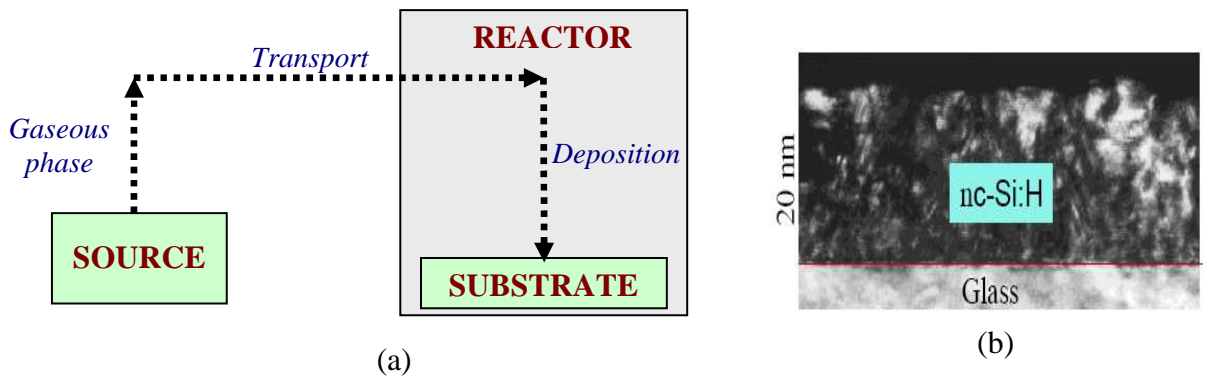

*Figure 2*: Principle of the Chemical Vapor Deposition (CVD) process:
*(a) Deposition and growth of Si solid phase from gaseous phase (typically $SiH_4$ and $H_2$ mixture)*
*(b) Example of 20 nm thickness nc-Si:H thin film deposited on glass substrate by CVD process.*

**III- Potential of nc-Si solar cells**

Nc-Si material presents many advantages [3, 4]. It is cheaper to produce than mono-crystalline silicon (c-Si), and thinner layers can be used to produce good solar cells with high efficiencies [11]. Also, it possess better electronic properties, higher stability, and practically no degradation after irradiation with light (figure 3) as frequently observed with a-Si:H [73-75]. In addition, due to its relatively lower energy band gap, it can be used in a tandem structure solar cell by combination with a-Si:H [12]. Because its grains are scaled down to nano-size and are accompanied by unique physical properties related to quantum size effect and Coulomb blockade phenomena, nc-Si material can also be applied to resonant tunneling devices, single electron transistors, and nano-memories [76]. Moreover, the lowering temperature of thin film deposition process, to enable performing devices on low cost substrates such as common glasses, flexible polymers, and stainless steel, became the main challenge to wide range of new microelectronic and optoelectronic applications. Thus, since it can be deposited at quite low temperatures and posses a great photo-electronic potential, nc-Si is presently considered among the very promising semiconductors for developing the next generation of thin film solar cells. This is why many research works on this material have been increasing over the last several years leading to various scientific publications and reports [3, 4].

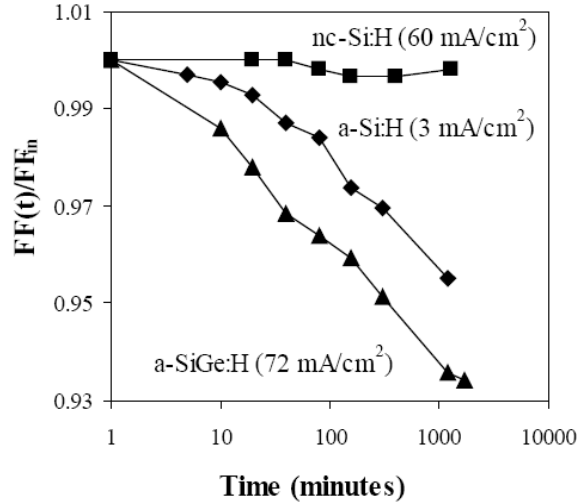

*Figure 3: Thin film solar cell fill factor (FF) soaking experiments. For nc-Si:H based photocell, no light-induced degradation was observed (after Uni-Solar Group)*

Although nc-Si cannot currently attain the mobility that conventional pc-Si can, it has the great advantage over the later one that it is easier to fabricate, as it can be deposited using common low temperature a-Si:H deposition techniques, such as plasma CVD, as opposed to laser annealing or high temperature processes, in the case of pc-Si. Moreover, since the remarkable differences of nc-Si from pc-Si are related to the presence of a high density of hydrogen atoms in the material, and because the presence of these atoms is a consequence to the low temperature deposition process, novel material science issues are also currently explored: the growth mechanism of crystalline silicon at temperatures much lower than the melting point, the role of hydrogen incorporated in silicon thin films and the effects of grain boundaries on the electrical properties of these films.

**IV- Low temperature process for nc-Si solar cells**

Nc-Si silicon solar cells are commonly operated in a p-i-n device structure because of the short minority carrier diffusion length ($L_D$), and the fact that photo-generated carriers are driven by the built-in electric field. Since the intrinsic layer (i-layer) contains defects density in the order of $10^{15-16}$ cm$^{-3}$, the band bending at the p-i and the i-n interfaces causes the upper limit of the i-layer thickness for the extraction of the photo-generated carriers. However, some experimental results suggest $L_D$ values over several hundreds nm, which are much larger than the evaluated grain size of the material. An anisotropic carrier transport has been also reported [77].

In solar cells, doped layers are crucial for all device performances. The p-type layer acts as a photocell window through which the light photons penetration occurs. Therefore, its electrical properties and optical transmittance in the short wavelengths region are both of great importance. The nc-Si p-type layer has advantages over hydrogenated amorphous silicon (a-Si:H) or hydrogenated amorphous silicon carbide (a-SiC:H) one in the respect of higher electrical conductivity and lower optical absorption. The p-type and n-type layers are basically formed by adding respectively diborane and phosphine gases as doping sources during the deposition process. However, the presence of these impurities while growing the layer influences the material crystallinity and structure. Indeed, intrinsic films show the best crystallinity around 400°C [3, 4] while B-doped films do so at around 140-180°C only. This difference is attributed to the catalytic role of boron to eliminate surface hydrogen [78]. So, with this result, the importance of the material surface coverage by hydrogen is once more supported. Therefore, it turns out that the low temperature growth of the p-type layer is effective to improve the crystallinity of nc-Si material. In addition, low temperature deposition is beneficial to prevent the band gap narrowing of the B-doped a-Si tissue region and the darkening of the textured Transparent Conductive Oxide (TCO) substrate due to the reduction by hydrogen atoms [79]. But, unfortunately, the electronic conductivity decreases significantly due to the hydrogen passivation of doping elements [80], and necessitates a post-annealing treatment to be restored [81]. Another important requirement is the reduction of the p-type layer optical absorption. To reach this aim, the film thickness should be minimized, while an appreciable quantity of doping atoms is required to form the built-in electric field, because there is a large distribution of dangling bonds that can act either as donor or acceptor defects in the i-layer. Therefore, both n-type and p-type layers should be heavily doped, while the p-layer is difficult to make crystalline as demonstrated by Fluckiger et al. who show that there is a difficulty during the crystalline nucleation for high boron concentration for thickness below 200 Å, which is a typical thickness of the p-layer [82].

In i-layer, the main impurities detected are oxygen, carbon and nitrogen. The well known oxygen-related donor defects, if activated, act as space charge and thereby screen the built-in potential in the layer. Therefore, the stronger electric field is formed at the p-i interface, as compared to the whole i-layer thickness. This results in the reduction of the active part of this thickness. It has been reported that the spectral response of the contaminated solar cells shows a marked deterioration in the long wavelengths region. This implies that the photo-generated holes cannot be extracted to the p-layer due to the weakened electric field (dead layer) in the deeper part of the i-layer. Even with oxygen contamination, it has been found that the low temperature deposition successfully passivates the oxygen-related donors in the films [3]. For example, in the case of the usual vacuum level (~ $10^{-5}$ Pa), in spite of the high incorporated oxygen density, which is more than $10^{19}$ cm$^{-3}$, the oxygen-related donors have been deactivated by reducing the deposition temperature to 140°C and, consequently, the performances of solar cells have been markedly improved [81]. The deposition temperature, however, is a very complicated parameter because it affects a variety of material properties such as cristallinity, defect density, interface damage, and so on. It has been established that the optimum temperature range is 140-180°C depending on other device parameters such as the i-layer thickness [83]. But, it is still an open question, however, whether a further improvement in solar cell efficiency is possible at higher temperature without oxygen contamination because the defect density of i-layer nc-Si shows a minimum at 200°C [84] similarly to a-Si:H.

Under fixed deposition conditions, the crystalline phase fraction in the film decreases with decreasing the temperature [85] and the solar cell efficiency decreases also [81]. However, this can be recovered by increasing the hydrogen dilution ratio during the deposition process. Under variable hydrogen dilution conditions for optimizing the solar cell efficiency, it turns out that this later can be maintained down to a deposition temperature of 120°C [81]. But, at this temperature regime, the hydrogen passivation of the doping elements becomes a more serious problem. Presently, the post annealing process is the only way to overcome this problem. The upper limit of the processing temperature, on the other hand, seems to be determined by the thermal damage at the p-i and n-i interfaces. The n-i-p solar cell can be successfully obtained over 300˚C [11], while the p-i-n cell is usually prepared at nearly 200˚C or lower [9, 86]. These results could arise from the different thermodynamics of the doping atoms.

Hydrogen dilution is an essential factor to determine the nc-Si based solar cell performances as well as the crystallinity of the material [87]. With increasing hydrogen dilution, the material structure varies from amorphous to crystalline and the crystalline volume fraction increases monotonically [87]. Therefore, one may expect the better performance for higher dilution ratio, whereas this is not the case. Indeed, while the open-circuit voltage ($V_{oc}$) monotonically increases with decreasing dilution ratio, the short-circuit current density ($J_{sc}$) still remains nearly constant in the crystalline regime and decreases abruptly in the amorphous regime. Therefore, an empirical conclusion is that the best performances are obtained near the amorphous-crystalline phase boundary in the crystalline regime.

## V- Understanding nc-Si growth

Normally, the crystalline phase fraction and grain size of nc-Si thin films determines its electronic and optical properties. During the deposition process, the nano-crystal nucleation mechanism, which dictates the final film structure, is governed by the interactions between the hydrogen atoms from the plasma and the solid silicon matrix growing on the substrate. Fundamental understanding of these interactions is important for optimizing the film structure and properties. In the following section, we briefly describe some of significant considerations about the growth chemistry of thin film a-Si:H material and how the hydrogen dilution influences its structural aspect and initiates the formation of nano-crystalline solid phase.

### V-1- Standard growth process

It is well known that during plasma deposition of a-Si:H thin films, a number of radicals are generated by the plasma inside the reactor. Among these are $SiH_3$, $SiH_2$, $SiH$ etc. In addition, hydrogen atoms are generated by decomposition of $SiH_4$, or of $H_2$ molecules if hydrogen is included as a diluent gas in the reactor. The standard growth model, first enunciated by Gallagher [88], states that the dominant radical leading to the growth of the a-Si:H films is $SiH_3$, because of its non reactivity with $SiH_4$, as opposed to the reactivity of $SiH_2$ with $SiH_4$. However, a smaller percentage of $SiH_2$ may be present in the reaction zone. Kushner estimated that the ratio of $SiH_2/SiH_3$ is about 10% [89]. Some others have come up with higher estimates of this ratio. Clearly, then, it is possible that some finite, non-negligible percentage of radicals present during growth is $SiH_2$, in addition to the dominant radical $SiH_3$.

The presence of more than one radical complicates the growth chemistry of a-Si:H films. Consider a silicon surface, with dangling bonds, waiting to accept an electron at each dangling bond site. If only $SiH_3$ radicals were present in the discharge, it would bond into this site, and give rise to a surface which was passivated (figure 4). From now on, it would not be possible to form another bond, and growth would cease. Clearly, then, the surface hydrogen has to be abstracted somehow. Two ways of doing this is with H atoms, forming $H_2$ molecules, or by $SiH_3$ radicals, forming $SiH_4$ molecules. Clearly, these two reactions have very different abstraction rates. Thus, even in the presence of only one radical in the plasma, a prescription exists for non-homogeneous growth.

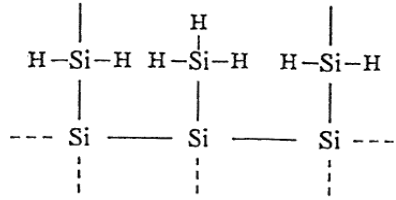

A non-homogeneous surface
in the absence of H dilution

*Figure 4: Schematic diagram showing how having different radicals present during deposition leads to the formation of both active and passive sites on the surface, leading to non-homogeneous growth.*

Now, if more than one radical were present, say $SiH_2$ and $SiH_3$, then the surface looks very different after insertion. One site may have $SiH_3$ inserted, with a resultant passive bond site, and another may have $SiH_2$ inserted, with a resultant active bond site (a site which can accept another radical). This is a prescription for very non-homogeneous growth, particularly at the low temperatures characteristic of a-Si:H growth.

Thus, it should come as no surprise that the standard a-Si:H is not homogeneous, and that its microstructure is full of voids, small and large [90]. And this microstructure depends on how one grows the material. As we change the material deposition conditions, we change the ion bombardment on the surface, we change the production and reaction rates of different radicals etc. As an example, if one were to use very low pressures, then $SiH_2$ would not have an opportunity to react with $SiH_4$, and therefore, the fraction of $SiH_2$ relative to $SiH_3$ may increase. Similarly, there would be little production of higher order $SiH_4$ molecules, such as $Si_2H_6$ at low pressures.

**V-2- Hydrogen dilution in the growth process**

A particularly interesting aspect is the dilution effect of hydrogen in the discharge. Experimentally, one finds that having significant dilution of hydrogen (10:1 or greater) helps in improving the stability of materials and devices [91]. The question is why. Considerations of growth chemistry provide the answer. At high hydrogen dilution, four effects take place: (1) the radical selection, (2) the surface homogenization, (3) the higher hydrogen radical flux induced homogenization and (4) the etching during growth.

*V-2-1- Radical selection*

In addition to the production of $SiH_3$ by electron impact of $SiH_4$ in the plasma, H can also produce $SiH_3$ through the reaction:

$$H + SiH_4 \Leftrightarrow H_2 + SiH_3$$

Thus, the presence of significant hydrogen dilution, when combined with the long mean free path of atomic hydrogen, can lead to a significant increase in the ratio of $SiH_3$ to $SiH_2$. This is entirely beneficial for film growth, because it reduces one aspect of non-homogeneous film growth. This mechanism is called radical selection.

*V-2-2- Surface homogenization*

During growth, at high dilution, the most likely reaction after radical insertion into an active bond is removal of surface hydrogen by atomic hydrogen. Thus, all sites will have their hydrogen removed by reacting with another hydrogen, as opposed to some sites having it removed by interacting with $SiH_3$ radical. Thus, the presence of a high density of atomic hydrogen results in homogenization of the surface, both in surface being bonded with hydrogen at all sites, independent of which radical inserted itself into the lattice, and in removal of this bonded hydrogen by another hydrogen (figure 5). This surface homogenization, in turn, promotes homogeneous growth and reduces clusters and voids, as is seen experimentally in post-growth studies. The hydrogen abstraction reaction by another hydrogen has been well proven by the work of Parsons et al. [92].

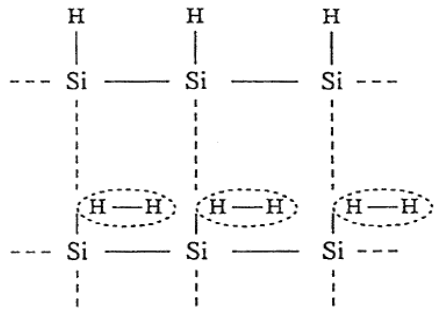

*Figure 5: Schematic diagram showing how the presence of significant hydrogen atoms during growth homogenizes the surface, and also allows for the abstraction of surface hydrogen by hydrogen atoms.*

*V-2-3- Hydrogen radical flux induced homogenization*

Since increasing the hydrogen dilution necessitates increasing the power in the discharge to achieve reasonable growth rates, the flux density of H radicals and ions impinging on the substrate also increases when hydrogen is used as a diluent. The increased ion density, and perhaps, increased ion energies, provide additional mobility to radicals on the surface, thereby promoting more homogeneous cluster-free growth.

*V-2-4- Etching process during growth*

Hydrogen, being highly reactive, etches the material during growth. Hydrogen penetrates a few layers deep into the material, and rearranges the bonds, etching away the poorly bonded Si atoms, thereby allowing for the growth of a more ordered structure. Indeed, in the extreme case, etching leads to crystallization of the film.

Thus, hydrogen plays very important and beneficial chemical and physical roles during deposition, and it should come as no surprise that a significant hydrogen dilution improves the stability of the material by turning it into nc-Si material.

## VI- Photo-deposition process for nc-Si solar cells

In most of the reports published so far on nc-Si material, the thin films have been grown under high radio frequency power density (> 100 mW/cm$^2$) using the technique of PECVD along with high hydrogen dilution [42-47]. Of course, H atoms are considered to be responsible for the silicon micro-crystallite formation through high surface coverage or etching action [28, 35]. But, the highly radio frequency power process generates highly energetic ions and electrons, which degrade interfaces and are also likely to damage substrates like the transparent conducting oxide (TCO) onto which the window layer of solar cells is grown. Such radiation induced damage from high energy charged particles may limit the conversion efficiency nc-Si based solar cells [93]. Hence there is a need to search for an alternative technique which allows good quality nc-Si thin film deposition at low temperature, and where energetic particle bombardments could be minimized or eliminated. Right now, these two conditions are fulfilled only by the Hg-Photo-CVD technique using high hydrogen dilution. Indeed, this process is rather soft since much lower power density (~ 10 mW/cm$^2$) of UV light is used to excite and dissociate the reactant gases. The UV radiations (185 nm and 253.7 nm) are expected to have an effect on the surface nucleation and hence assist the formation of silicon nano-crystallites [94] since the photon energies (6.71 and 4.89 eV) are greater than the barrier height of silicon crystallization (4.2 eV). The photon energy is not high enough to ionize the reactant species. Thus, in absence of electrons and ions, photo-CVD is expected to produce ion-damage-free high quality thin film silicon material. Recently, we have published a paper in which we provided an overview on this technique [95, 96] and now we propose to provide the reader with an up-dated version.

## VII- Technique of Photo-Chemical Vapor Deposition

Photo-Chemical Vapor Deposition (Photo-CVD) technique, also named Photo-Enhanced-CVD or quite simply PVD [95-103] is widely used for growing thin silicon films. As only atomization and no ionization of the gas-phase reactants is involved in the photo-deposition process, this attractive method allows damage-free thin film growth at very low substrate temperatures (~ 100$^o$C) without the deleterious effects of the other various techniques [104, 105]. Because of the efficient generation of both SiH$_3$ radical, having a long lifetime in the gas-phase, and atomic hydrogen, having an important role of the surface termination, high quality amorphous, micro-crystalline, nano-crystalline and epitaxial silicon films were obtained by this method [106].

As it is known, there are two major variants of the Photo-CVD technique depending on the source of excitation used in the decomposition process:

- laser-induced Photo-CVD, which use high-energy coherent radiations as light source to cause a direct photolysis mechanism [107-109].

- lamp-induced Photo-CVD, which use incoherent radiations as light source to perform an indirect photolysis mechanism. A few mercury atoms (Hg) are usually needed as photo-sensitizers for the deposition process (Hg-sensitization) because of mercury's high catalytic activity [50-52]. This is why such technique is named Hg-sensitized photo-CVD (figure 6).

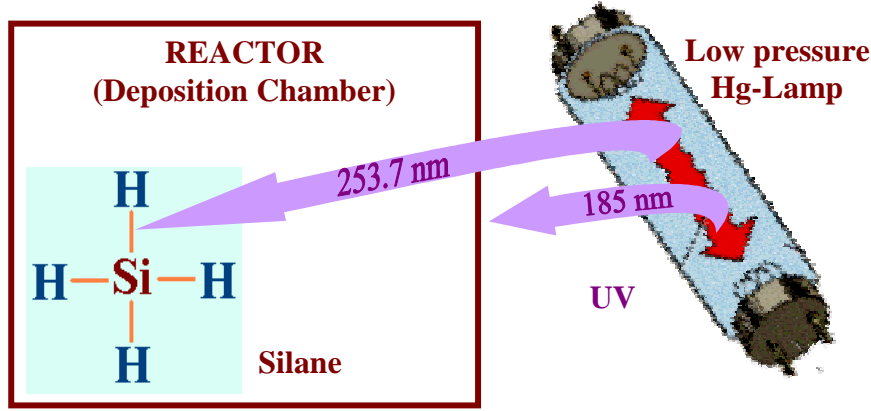

*Figure 6: Principle of Hg-sensitized Photo-CVD.*
*To break the Si-H bonds, $SiH_4$ molecules must absorb the UV radiation $\lambda$ = 253.7 nm.*

In the first variant, various gas lasers such as excimers [97, 107], $Ar^+$ ions [108], and $CO_2$ lasers [109] are often used because of their high power density. However, they are expensive and are inadequate as vacuum-ultra-violet (VUV) light sources. Moreover, a strict optimization of the technology is necessary. Namely, some effects, such as damage due to the high-energy photons may have disadvantageous results. This is why the second variant Hg-Photo-CVD, which operates in a indirect photo-CVD mode without the use of lasers, has been widely utilized for practical applications in spite of its principal disadvantages such as Hg incorporation in the growth films [110] and the production of toxic exhaust fumes.

**VII-1- Hg-sensitized photo-CVD system**

The schematic diagram of the photo-CVD apparatus we are developing in our laboratory was described elsewhere [95, 96]. The set-up is very simple (figure 7). It consists mainly of a cylindrical horizontal quartz reactor in the form of special tube whose size dimensions as specified in figure 8. This tube, used as deposition chamber, is equipped with a small reservoir containing a small quantity of liquid mercury, which can be thermally controlled independently of the remainder of the system (Hg bath). The inner surface of the tube is coated with a low-vapor-pressure Fomblin vacuum oil to prevent any film deposition on. The substrate-support can be of graphite or of stainless steel. It is heated through a thermo-coax wire and equipped with a thermocouple to measure its temperature. Well-cleaned high resistivity crystalline Si (100) wafers or Corning 7059 glass can be used as substrates. The distance between the quartz tube external surface and substrates is about 2 cm. The UV light source consists of a series of low pressure Hg lamps radiating both 253.7 nm (40 mW/cm$^2$ or ~ 30 mW/cm$^2$ at 3 cm distance) and 184.9 nm (less than 10 mW/cm$^2$ or ~ 5 mW/cm$^2$ at 3 cm distance) resonance lines. Since the transmittance of the quartz tube for the 253.7 nm and 184.9 nm wavelengths is 80 % and 20 % respectively, the 253.7 nm resonance line of UV light is dominantly irradiated into the reactor. The series of Hg lamps are set up under an aluminum reflector placed at few cm with the top of the reactor. The vacuum system is composed of two pumps. The first one, with pallets, is used to obtain a primary vacuum and to purge gases of the reactor. The second is a diffusion pump allowing evacuation down to a pressure of $10^{-6}$ Pa during the back out of the reactor prior to the growth. The reactant gases ($SiH_4$, $H_2$, …) are introduced into the reactor through the Hg bath usually kept at temperature between 35 and 50ºC. Thus, a very small amount of Hg vapor is automatically mixed with the gases and introduced into the deposition chamber. The Hg vapor atoms then introduced are used to enhance dissociation of reactant gases [111] because of the weak optical absorption by the molecules in the 190-200 nm wavelengths region [112, 113]. A conductance valve adjusts the total gas pressure in the reactor.

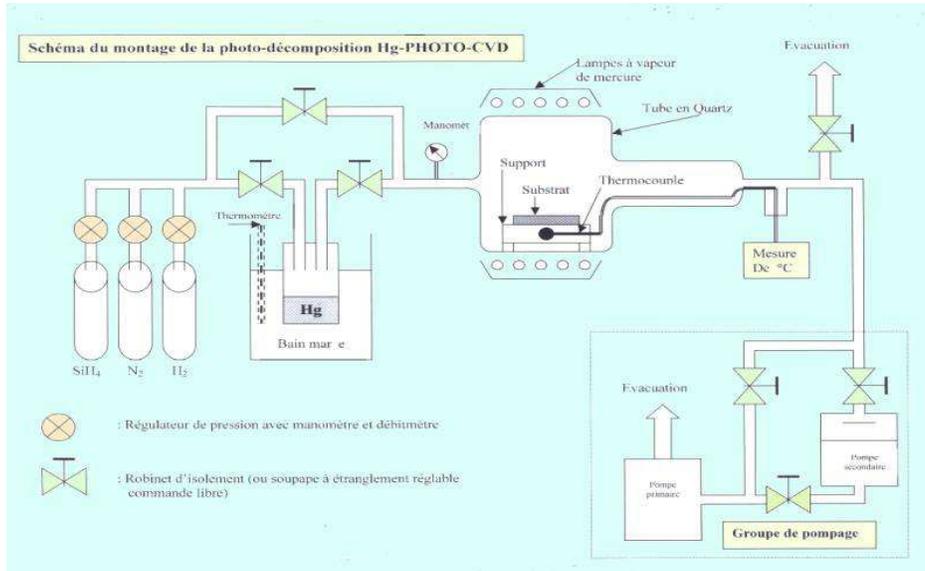

*Figure 7: Schematic diagram of the Hg-sensitised photo-chemical vapor deposition apparatus developed at home laboratory*

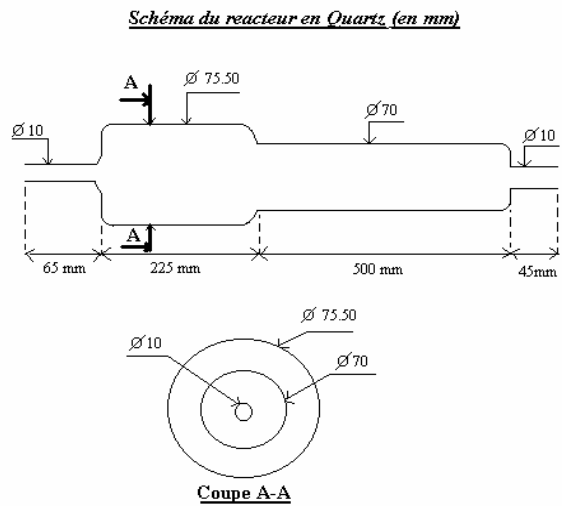

*Figure 8: Schema and size dimensions of the cylindrical horizontal quartz reactor used for the Hg-sensitized photo-chemical vapor deposition apparatus we are developing at home laboratory*

## VII-2- Principle of Hg-sensitized photo-CVD

As well described in the literature [114, 115], in the principle of Hg-sensitized photo-CVD technique, the 253.7 nm resonant radiation emitted by the low pressure Hg-lamps is used as UV light source. The photons excite the Hg atoms from the ground electronic state $Hg^o$ ($\equiv Hg\ (^1S_0)$) to the excited electronic state $Hg^*$($\equiv Hg\ (^3P_1)$). Then the $SiH_4$ molecules are decomposed by colliding with the excited Hg atoms [116]. Thus, the photochemical reactions are much more selective than the reactions in the case of plasma-CVD. As a result of the Hg-sensitized reactions of $SiH_4$, the radicals such as $SiH_3$ and H single atoms are generated as described in the reactions 1 and 2 below. $H_2$ molecules are also decomposed by the excited Hg and yield atomic H (reaction 3) [117].

$$Hg^o\ (\equiv Hg\ (^1S_0)) + h\nu\ (\lambda = 253.7\ nm) \longrightarrow Hg^*\ (\equiv Hg\ (^3P_1)) \quad (1)$$
$$Hg^*(\equiv Hg\ (^3P_1)) + SiH_4 \longrightarrow SiH_3 + H + Hg^o\ (\equiv Hg\ (^1S_0)) \quad (2)$$
$$Hg^*(\equiv Hg\ (^3P_1)) + H_2 \longrightarrow 2H + Hg^o\ (\equiv Hg\ (^1S_0)) \quad (3)$$

To calculate the concentrations of $SiH_3$ radicals and H atoms, an excited Hg ($Hg^*$) concentration inside the reaction chamber must be calculated. The method for the calculation has been described by Perrin et al. [118].

## VII-3- Analysis of gas-phase reaction

Once the $Hg^*$ concentration in the reaction chamber is calculated, the concentrations of the radicals generated by chemical reactions can also be estimated. The details of the calculation used to determine the radical concentrations are described elsewhere [115]. The elemental gas-phase reactions basically and usually adopted are as follows [119-122]:

$$h\nu\ (\lambda = 253.7\ nm) + Hg\ (^1S_0) \longrightarrow Hg^*\ (\equiv Hg\ (^3P_1)) \quad (4)$$
$$Hg^* + SiH_4 \longrightarrow Hg\ (^1S_0) + H + SiH_3 \quad \text{(rate constant } k_1) \quad (5)$$
$$H + SiH_4 \longrightarrow H_2 + SiH_3 \quad \text{(rate constant } k_2) \quad (6)$$
$$SiH_3 + SiH_3 \longrightarrow SiH_2 + SiH_4 \quad \text{(rate constant } k_3) \quad (7)$$
$$SiH_2 + SiH_4 \longrightarrow Si_2H_6 \quad \text{(rate constant } k_4) \quad (8)$$
$$H_2 + Hg^* \longrightarrow Hg\ (^1S_0) + 2H \quad \text{(rate constant } k_5) \quad (9)$$
$$H + SiH_3 \longrightarrow H_2 + SiH_2 \quad \text{(rate constant } k_6) \quad (10)$$
$$SiH_2 + H_2 \longrightarrow H + SiH_3 \quad \text{(rate constant } k_7) \quad (11)$$

The rate constants $k_i$ at 300K in $cm^3\ mol^{-1}\ s^{-1}$ are as follows:

$$k_1 = 6.4\ 10^{-10}$$
$$k_2 = 4.3\ 10^{-13}$$
$$k_3 = 1.5\ 10^{-10}$$
$$k_4 = 1.0\ 10^{-12}$$
$$k_5 = 5.0\ 10^{-10}$$
$$k_6 = 1.0\ 10^{-10}$$
$$k_7 = 2.6\ 10^{-13}$$

The radical-radical reaction of hydrogen (H + H → H$_2$) is ignored in the calculation because of the very small rate constant of 4.3 10$^{-16}$ cm$^3$ mol$^{-1}$ s$^{-1}$ [123]. Consequently, the radical concentrations are calculated by solving the following gas-phase mass balance equations of SiH$_3$, SiH$_2$ and H with the boundary conditions on the Si substrate and the inner front surface of the quartz tube wall from which reactant gases in the reactor receive the UV light irradiation. In these calculations, we assume that the UV light is irradiated uniformly and that the macroscopic flow of the reactant gas is negligible:

$$-D_{SiH_3}\frac{d^2[SiH_3]}{dx^2} = k_1[SiH_4][Hg^*] + k_2[SiH_4][H] - 2k_3[SiH_3][SiH_3] - k_6[SiH_3][H] + k_7[SiH_2][H_2] \quad (12)$$

$$-D_{SiH_2}\frac{d^2[SiH_2]}{dx^2} = k_3[SiH_3][SiH_3] - k_4[SiH_2][SiH_4] + k_6[SiH_3][H] - k_7[SiH_2][H_2] \quad (13)$$

$$D_H\frac{d^2[H]}{dx^2} = k_1[SiH_4][Hg^*] - k_2[SiH_4][H] + 2k_5[H_2][Hg^*] - k_6[SiH_3][H] + k_7[SiH_2][H_2] \quad (14)$$

The boundary conditions for these equations are shown in equations (15) and (16), that is, the flux of each species impinging onto the surface is equal to the reaction rate at the surface:

$$\left[D_j\frac{dn_j}{dx}\right]_{x=0} = \left[D_j\frac{\beta_j \upsilon_j n_j}{4}\right]_{x=0} \quad (15)$$

$$-\left[D_j\frac{dn_j}{dx}\right]_{x=L} = \left[D_j\frac{\beta_j \upsilon_j n_j}{4}\right]_{x=L} \quad (16)$$

where x is the distance between the substrate and the inner front surface of the quartz tube wall from which reactant gases in the reaction chamber receive the UV light irradiation, $D_j$ is a diffusion coefficient of the radical j (SiH$_3$, SiH$_2$ or H) in the gas phase and $n_j$ and $\beta_j$ are the concentration and the reaction probability ($0 \leq \beta_j \leq 1$) respectively. The man velocity $\upsilon_j$ is given by $(8\,R\,T_g\,/\,\pi\,M_j)^{1/2}$ where $M_j$ is the mass of the radical j.

The values of $\beta_j$ on Fomblin oil (x = 0) and on the surface of Si films (x = L) are usually adopted with the following assumptions: On Fomblin oil, $\beta_{SiH_3}$ and $\beta_{SiH_2}$ are smaller than 10$^{-3}$ since the growth rate of Si films on Fomblin oil is very small. The value of $\beta_H$ on Fomblin oil has been reported to be about 10$^{-1}$ [124]. On the Si surface, SiH$_2$ has a high sticking coefficient, since SiH$_2$ can insert in a Si-H bond, therefore $\beta_{SiH_2}$ of 0.7 has been selected in previous study [115]. $\beta_{SiH_3}$ on Si film was determined by Perrin [125] and has been reported to be 10$^{-1}$. The reported values of $\beta_H$ ranged from 0.6 to 0.8.

**VII-4- Surface-reaction model of SiH$_3$ and H**

The surface-reaction model of SiH$_3$ and H has been well described by Abe et al. [114]. At temperatures lower than 250°C, the Si(100) surface is assumed to be almost terminated by hydrogen atoms with forming di-hydride structures [126]. The radicals impinging on the surface migrate and react on the H-terminated surface. As a surface-reaction model, one always considers only two species, SiH$_3$ and H, and two types of sites on the surface, which are dangling bonds (DB) and Si-H bonds (-H). The effects of SiH$_2$ radicals are neglected because the flux ratio SiH$_2$ / SiH$_3$ on the surface is estimated to be about 0.01 under experimental conditions [114]. The following three reactions are taken into account for the reaction model of silicon epitaxy on the di-hydride surface:

| | | | | |
|---|---|---|---|---|
| SiH$_3$ + DB + (-H) | $\longrightarrow$ | =SiH$_2$ + H$_2$ | : film growth | (17) |
| SiH$_3$ + (-H) | $\longrightarrow$ | SiH$_4$ | : H abstraction | (18) |
| H + DB | $\longrightarrow$ | (-H) | : H termination | (19) |

The first reaction is the film growth: SiH$_3$ radical migrates on the H-terminated Si (100) surface, and then the radical is chemisorbed at dangling bond, resulting in the growth of Si. Here, "=SiH$_2$" is a newly created di-hydride structure. The second reaction is the abstraction of bonding hydrogen by SiH$_3$ radical. The third reaction is the dangling bond termination by atomic hydrogen. Therefore, from this surface-reaction model, the equation for the change in the hydrogen surface-coverage ratio "θ" is given by:

$$N_S (d\theta/dt) = \beta_S J_S N_S (1-\theta) L_S \sigma_A - \beta_S J_S N_S \theta L_S \sigma_B + \beta_H J_H N_S (1-\theta) L_H \sigma_C \quad (20)$$

where L$_S$ and L$_H$ are the migration lengths of SiH$_3$ and H respectively. σ is a two dimensional cross section of each reaction. ß is the total surface-loss probability of each radical, and the values of ß$_S$ and ß$_H$ are reported to be 0.1 and 0.8 respectively [125]. J$_S$ and J$_H$ are the flux of SiH$_3$ and H, which can be obtained through calculation of the gas-phase reactions, respectively. N$_S$ (6.8 10$^{14}$ cm$^{-2}$) is the surface atomic density of Si (100).

In steady state, (dθ/dt) goes to zero. Thus, equation (20) is deduced to be:

$$\theta = \frac{\beta_H J_H L_H \sigma_C + \beta_S J_S L_S \sigma_A}{\beta_H J_H L_H \sigma_C + \beta_S J_S L_S (\sigma_A + \sigma_B)} \quad (21)$$

From the growth reaction, the growth rate equation is also given by:

$$\text{Growth rate} = \beta_S J_S N_S (1-\theta) L_S \sigma_A \, m / \rho \quad (22)$$

where m (4.66 10$^{-23}$ g) is the mass of Si atom and ρ (2.329 g/cm$^3$) is its atomic density.

Using equation (21), equation (22) is transformed into the following:

$$\text{Growth rate} = \frac{(\beta_S J_S)^2 L_S \sigma_A L_S \sigma_B N_S m / \rho}{\beta_H J_H L_H \sigma_C + \beta_S J_S (L_S \sigma_A + L_S \sigma_B)} \quad (23)$$

## **VIII- Kinetic model for lamp-induced Photo-CVD**

This kinetic model, based on collision theory of chemical reactions, has been proposed for laser and lamp-induced photo-CVD from GeH$_4$ and SiH$_4$ by Tao [127]. In a lamp-induced photo-CVD reactor, some reactant molecules absorb the UV photons and are photo-chemically excited. Their deposition is classified as photochemical deposition. Most reactant molecules do not have the chance to absorb the UV photons, so their deposition is still thermally driven. The overall deposition is composed of photochemical deposition and thermal deposition.

In an elementary gas-solid deposition reaction, there are two premises for a reactant molecule to deposit onto the substrate, according to the collision theory of chemical reactions [114]: (1) The reactant molecule must have enough energy to be activated and (2) the reactant molecule must collide with the substrate to deposit onto it. The growth rate depends on the number of activated reactants molecules, which strike the substrate.

Activation means that there is threshold energy $E_c$. If the energy of a reactant molecule is higher than $E_c$, it will decompose and deposit when it strikes the substrate. Otherwise, it will return into the gas phase after a collision. In the collision theory, a molecule is treated as a hard sphere, so its energy consists of its kinetic energy $E_k$, its potential energy $E_p$, and its internal energy $E_i$. Activation requires that:

$$E_k + E_p + E_i \geq E_c \qquad (24)$$

For an ideal gas, the potential energy $E_p = 0$. In the classical thermal deposition process, reactant molecules are not photo-excited. Their internal energies can be approximated by a constant, which is the ground-state energy $E_i^0$. Thus, Equation 24 can be simplified as:

$$E_k \geq E_c - E_i^0 = E_a \qquad (25)$$

Accordingly, activation means that the kinetic energy of a reactant molecule must be higher than a critical value $E_c - E_i^0$, denoted by $E_a$.

In lamp-induced photochemical deposition, some reactant molecules absorb the UV non-monochromatic photons emitted by the lamps and are photo-chemically excited. The excitation is either dissociate or non-dissociate. In the non-dissociate case, the internal energies of photo-excited molecules are no longer at the ground state:

$$E_i = E_i^0 + \sum_j h\nu_j \qquad (26)$$

where $h\nu_j$ is the energy of a particular photon, h is the Planck constant, $\nu_j$ is the frequency of the photon and the summation is over all the photons absorbed by a reactant molecule.

Thus, equation 26 can be written as:

$$E_k \geq E_c - E_i^0 - \sum_j h\nu_j = E_a - \sum_j h\nu_j = E_a^* \qquad (27)$$

which represents the lamp-induced photo-chemical excitation process.

From this equation, we can easily note that the apparent activation energy becomes smaller in non-dissociate photochemical deposition than in the classical thermal deposition [128].

In the dissociate case, photo-excited molecules turn into new species. Equation 24 can be applied to the new species:

$$E_k \geq E_a^* \tag{28}$$

where $E_a^*$ is the activation energy for new species to decompose and deposit onto the substrate.

As discussed above, the activation energy for new species deposition is smaller than that of the parent molecules.

At equilibrium, the Maxwell distribution [129] indicates that, in a unit time, the number of reactant molecules which collide with a unit area of the substrate with kinetic energies between $E_k$ and $E_k + dE_k$ is:

$$d\Gamma = 2\pi\gamma N_r \left(\frac{1}{2\pi m_r^* k T}\right)^{3/2} m_r^* E_k \exp\left(-\frac{E_k}{kT}\right) dE_k \tag{29}$$

where $N_r$ is the number of reactant molecules in a unit volume of the gas phase, $\gamma$ is the fraction of those photo-excited.

So, the number of photo-excited reactant molecules in a unit volume of the gas phase is given by $N^* = \gamma N_r$. $m_r^*$ is the mass of a photo-produced species, k is the Boltzmann constant, and T is the absolute temperature.

Integrating equation 29 from $E_a^*$ to $+\infty$, we can derive the number of the activated photo-excited reactant molecules which strike a unit area of the substrate in a unit time:

$$\Gamma = \gamma N_r \left(\frac{1}{2\pi m_r^* k T}\right)^{1/2} (E_a^* + kT) \exp\left(-\frac{E_a^*}{kT}\right) \tag{30}$$

We have the ideal gas law $P_r = N_r k T$, so the growth rate from photo-excited molecules is:

$$R_p = \frac{\Gamma}{N_0} = \gamma \frac{P_r}{N_0} \left(\frac{1}{2\pi m_r^* k T}\right)^{1/2} \left(\frac{E_a^*}{kT} + 1\right) \exp\left(-\frac{E_a^*}{kT}\right) \tag{31}$$

where $P_r$ is the partial pressure of the photo-excited reactant molecules, $N_0$ is the number density of atoms in the growing film.

In the non-dissociate case, $m_r^* = m_r$. Simultaneously, thermal deposition proceeds as:

$$R_t = (1-\gamma)\frac{P_r}{N_0}\left(\frac{1}{2\pi m_r kT}\right)^{1/2}\left(\frac{E_a}{kT}+1\right)\exp\left(-\frac{E_a}{kT}\right) \quad (32)$$

The total growth rate is then given by the following equation 33:

$$R = R_p + R_t = \gamma\frac{P_r}{N_0}\left(\frac{1}{2\pi m_r^* kT}\right)^{1/2}\left(\frac{E_a^*}{kT}+1\right)\exp\left(-\frac{E_a^*}{kT}\right) + (1-\gamma)\frac{P_r}{N_0}\left(\frac{1}{2\pi m_r kT}\right)^{1/2}\left(\frac{E_a}{kT}+1\right)\exp\left(-\frac{E_a}{kT}\right) \quad (33)$$

It should be noted that this last equation is valid for small γ, which is the fraction of photo-excited molecules. Only when γ is small can we expect a simple linear behavior. When γ is large, gas-phase reactions can be very complicated.

A common case in lamp-induced photo-CVD is that a photo-sensitizer, usually mercury, is used to absorb photon energy and then transfer the energy to reactant molecules by collision interactions (Hg-sensitized photo-CVD). The chemistry and physics in this case are more complicated than in direct photochemical excitation. However, the result of the indirect photochemical excitation is similar to that of the direct one: $SiH_4$ molecules are converted into more reactive species, such as $SiH_2$, $SiH_3$, $Si_2H_6$, $Si_2H_5$ and $Si_2H_4$ [130]. Equation 28 can be applied to these new species, and the total growth rate has the same format as equation 33.

In the literature, the experimental growth rate of Hg-sensitized photo-CVD from $SiH_4$ as a function of Hg vaporizer temperature [131] has been compared with this present model. The fraction of photo-excited molecules γ is, in the simplest case, proportional to the partial pressure of Hg, which can in turn be expressed as:

$$P_{Hg} = A\exp\left(-\frac{\Delta H_v}{kT_{Hg}}\right) \quad (34)$$

where $P_{Hg}$ is the partial pressure of Hg, A is the pre-exponential factor, $\Delta H_v$ is the heat of vaporization, and $T_{Hg}$ is the temperature of the Hg vaporizer.

Therefore, γ can be written in the same way with a new pre-exponential factor B:

$$\gamma = B\exp\left(-\frac{\Delta H_v}{kT_{Hg}}\right) \quad (35)$$

Combining Equation 31 and Equation 35, we get the growth rate from photo-excited molecules as a function of Hg vaporizer temperature:

$$R_p = B\exp\left(-\frac{\Delta H_v}{kT_{Hg}}\right)\frac{P_r}{N_0}\left(\frac{1}{2\pi m_r^* kT}\right)^{1/2}\left(\frac{E_a^*}{kT}+1\right)\exp\left(-\frac{E_a^*}{kT}\right) \quad (36)$$

In summary, lamp radiation produces photo-excited molecules, which are more reactive than parent molecules. The photochemical excitation of reactant molecules is either dissociate or non-dissociate. Lamp-induced photo-CVD is composed of two processes: extrinsic photochemical deposition from photo-excited molecules and intrinsic thermal deposition from normal reactant molecules. The growth rates of both photochemical deposition and thermal deposition are derived by means of statistical physics. Photochemical deposition dominates at low temperatures, and thermal deposition becomes prominent as the temperature rises. The transition temperature from photochemical deposition to thermal deposition is obtained as a function of fraction of photo-excited molecules and activation energies of photochemical deposition and thermal deposition.

## IX- Conclusion

Photochemical Vapor Deposition (Photo-CVD) technique is widely used for growing thin silicon films. This attractive method allows damage-free thin film depositions at very low substrate temperatures without the deleterious effects of the other various techniques. After examining and analyzing various works published in the literature on Photo-CVD process and experimental procedures, we noted some fundamental aspects, which characterize particularly the Mercury-sensitized photo-CVD technique. This review reports on the the great photovoltaic potential of this technique to perform low temperature deposition of nano-crystalline silicon thin films for the development of the new generation of high efficiency low cost solar cells

## Acknowledgments

This paper has been prepared on the basis of bibliographical investigation and analysis. The author would like to thank Professor G. Furlan, Head of ICTP-TRIL Program, for his kind co-operation and paper referring. The author thanks are also addressed to his colleagues at home laboratory Profs. N. M'Gafad and S. Karbal for their precious contributions and fruitful discussions, and to his collaborators Prof. M. C. El-Idrissi and Ing. A. Benmakhlouf from "Physique des Surfaces et Interfaces" laboratory, Faculty of Sciences, Ibn Tofail University, Morocco, for their kind and efficient technical help.

## References

[1]  M. A. Green
     Third Generation Photovoltaics: Advanced Solar Energy Conversion
     Springer Series in Photonics, N°5, 2003
[2]  A. Goetzberger and V. U. Hoffmann
     Photovoltaic Solar Energy Generation
     Springer Series in Optical Sciences. N°112, 2005
[3]  M. Kondo and A. Matsuda
     In "Thin Film Solar Cells: Next Generation Photovoltaics and Applications"
     Chap. 4, pp. 69-89.
     Editor: Y. Hamakawa, 2004, Springer Series in Photonics

[4]   M. Kondo and A. Matsuda
      In "Thin Film Solar Cells: Next Generation Photovoltaics and Applications"
      Chap. 8, pp. 138-149
      Editor: Y. Hamakawa, 2004, Springer Series in Photonics
[5]   A. Shah, P. Torres, R. Tscharner, N. Wyrsch and H. Keppner
      Science, 285, 1999, pp. 692-698.
[6]   R. A. Street
      Phys. Status Solidi A, 166, 1998, pp. 695-705.
[7]   S. Veprek and V. Marecek
      Solid State Electron., 11, 1968, p. 683.
[8]   R. B. Bergmann
      Appl. Phys. A, 69, 1999, p. 155
[9]   J. Meier et al.
      Proc. Mater. Res. Soc. Symp., 420, 1996, p. 3.
[10]  K. Yamamoto, T. Suzuki, M. Yochimi and A. Nakajima
      Jpn. J. Appl. Phys., 36, 1997, p. L569.
[11]  K. Yamamoto, M. Yochimi, Y. Tawada, K. Okamoto, A. Nakajima and S. Igari
      Appl. Phys. A, 69, 1999, p. 179 and references therein.
[12]  K. Yamamoto
      Tech. Digest of PVSEC 12, Jeju, korea, 2001, p. 547.
[13]  A. Madan
      MVSystems, Inc., Golden, Colorado
      Final Technical Report, NREL/SR-520-37718, Marsh 2005.
[14]  S. Sivoththaman, B. Hartiti, J. Nijs, A. Barhdadi, M. Rodot, J-C. Muller, W. Laureys, D. Sarti
      Proc. of the 12th European Photovoltaic Solar Energy Conference and Exhibition,
      11 - 15 April 1994, Amsterdam, Netherlands. pp. 47-51.
[15]  J-C. Muller, P. Siffert, A. Barhdadi and H. Amzil
      J. Chim. Phys., 88 (10), 1991, pp. 2223-2228.
[16]  J-C. Muller, Y. Ababou, A. Barhdadi, E. Courcelle, S. Unamuno, D. Salles, J. Fally and P. Siffert
      Solar Cells, 17, 1986, pp. 201-231.
[17]  A. Barhdadi, J-P. Ponpon, A. Grob, J-J. Grob, A. Mesli, J-C. Muller and P. Siffert
      Vacuum, 36 (10), 1986, pp. 705-709.
[18]  A. Slaoui, A. Barhdadi, J-C. Muller and P. Siffert
      Appl. Phys. A, 39, 1986, pp. 159-162.
[19]  J-C. Muller, A. Barhdadi, Y. Ababou and P. Siffert
      Rev. Phys. Appl., 22, 1987, pp. 649-654.
[20]  A. Barhdadi, N. M'gafad, H. Amzil, J-C. Muller, V-T. Quat and P. Siffert
      Proc. of the 8th European Photovoltaic Solar Energy Conference,
      9 - 13 May 1988, Florence, Italy, Klumer Academic Publishers, Vol. 2, pp. 1427-1431.
[21]  J-C. Muller, V-T. Quat, P. Siffert, H. Amzil, A. Barhdadi and N. M'gafad
      Solar Cells, 25, 1988, pp. 109-125.
[22]  A. Barhdadi, H. Amzil, J-C. Muller and P. Siffert
      Springer Proceedings in Physics, Vol. 35 (1989) pp. 158-163,
      Springer-Verlag Berlin, Heidelberg, Ed.: H-J. Moller, H-P. Strunk and J-H. Werner.

[23] W. E. Spear and P. G. LeComber
    Solid State Commun., 17, 1975, p. 1193.
[24] K. Nomoto, Y. Urano, J. L. Guizot, G. Ganguly and A. Matsuda
    Jpn J. Appl. Phys., 2 (29), 1990, pp. L1372-L1375.
[25] M. Katiyar, and J. R. Abelson
    Mater. Sci. Eng., A 304, 2001, pp. 349-352.
[26] N. Layadi, P. Roca i Cabarrocas, B. Drevillon and I. Solomon
    Phys. Rev. B, 52, 1995, pp. 5136-5143.
[27] A. Fontcuberta i Morral, J. Bertomeu and P. Roca i Cabarrocas
    Mater. Sci. Eng., B 69, 2000, pp. 559-563.
[28] C. C. Tsai, G. B. Anderson, R. Thompson and B. Wacker
    J. of Non-Cryst. Solids, 114, 1989, pp. 151-153.
[29] J. R. Abelson
    Appl. Phys. A, 56, 1993, pp. 493-512.
[30] K. Saitoh et al.
    Appl. Phys. Lett., 71, 1997, pp. 3403-3405.
[31] C. Summonte et al.
    Phil. Mag. B, 80, 2000, pp. 459-473.
[32] L. Kaiser, N. H. Nickel, W. Fuhs and W. Pilz
    Phys. Rev. B, 58, 1998, pp. R1718-R1721.
[33] J. J. Boland and G. N. Parsons
    Science, 256, 1992, pp. 1304-1306.
[34] N. H. Nickel and W. B. Jackson
    Phys. Rev. B, 51, 1995, pp. 4872-4881.
[35] H. Shirai, J. Hanna and I. Shimizu
    Jpn. J. Appl. Phys., 30, 1991, pp. L679-L682.
[36] S. Sriraman, S. Agarwal, E. S. Aydil and D. Maroudas
    Nature, 418, 2002, pp. 62-65.
[37] Y. L. Wang, G. S. Fu, Y. C. Peng, Y. Zhou, L. Z. Chu and R. M. Zhang
    Chin. Phys. Lett., 21 (1), 2004, pp. 201-202
[38] G. S. Fu, W. Yu, S. Q. Li, H. H. Hou, Y. C. Peng and L. Han
    Chin. Phys., 12 (1), 2003, pp. 75-78
[39] W. Yu, B. Z. Wang, W. B. Lu, Y. B. Yang, L. Han and G. S. Fu
    Chin. Phys. Lett., 21 (7), 2004, pp. 1320-1322
[40] M. Morales, Y. Leconte, R. Rizk and D. Chateigner
    J. Appl. Phys., 97 (3), 2005, pp. 4307-4320
[41] H. R. Moutinho, C. S. Jiang, Y. Xu, B. To, K. M. Jones, C. W. Teplin, M. M. Al-Jassim
    Proc. of the 31st IEEE Photovoltaics Specialists Conference and Exhibition
    Lake Buena Vista, Florida, January 3-7, 2005
[42] P. Roca i Cabarrocas
    Phys. Stat. Sol. C, 1 (5), 2004, pp. 1115-1130
[43] J. Müller, F. Finger, R. Cariuus and H. Wagner
    Phys. Rev. B, 60, 1999, p. 11666.
[44] U. Kroll, J. Meier, A. Shah, S. Mikhailov and J. Weber
    J. appl. Phys., 80, 1996, p. 4971.

[45] M. Kondo, Y. Nasumo, H. Mase, T. Wada and A. Matsuda
J. Non-Crys. Solids, 299-302, 2002, p. 108.
[46] S. Hamma, P. Roca i Cabarrocas
J. Non-Crys. Solids, 227-230, 1998, p. 852.
[47] P. Hapke and F. Finger
J. Non-Crys. Solids, 227-230, 1998, p. 861.
[48] K. Endo et al.
Solar Energy Materials & Solar Cells, 66, 2001, p. 283.
[49] S. Y. Myong, T. H. Kim, K. S. Lim, K. H. Kim, B. T. Ahn, S. Miyajima and M. Konagai
Solar Energy Materials & Solar Cells, 81, 2004, pp. 485-493.
[50] Y. K. Su, C. J. Wang and Y. C. Chou
Jpn. J. Appl. Phys. 28 (1989) 1644
[51] C. J. Hwang and Y. K. Su
J. Electron. Mater. 19 (1990) 753.
[52] C. J. Hwang and Y. K. Su
J. Appl. Phys. 67 (1990) 3350
[53] B. Aka,
Doctorate thesis, Université of Louis Pasteur, Strasbourg I, 1989, France.
[54] B. Aka,
J. Soc. Ouest Afr. Chim., 10 (2001) 119
[55] B. Aka, G.A. Monnehan, C. Fuchs and E. Fogarassy
Rev. Iv. Sci. Tech., 2 (2001) 49
[56] B. Aka and A. Trokourey,
J. Soc. Ouest Afr. Chim., 9 (2000) 27
[57] S. Oda
Nanonet Interview, Japan Nanonet Bulletin, 8$^{th}$ Issue, December 25, 2003
[58] E. Edelberg, S. Bergh, R. Naone, M. Hall and E. S. Aydil
J. Appl. Phys., 81, 1997, pp. 2410-2417.
[59] P. Roca i Cabarrocas
J. Non-Cryst. Solids, 266, 2000, pp. 31-37.
[60] B. C. Pan and R. Biswas
J. of Non-Crystalline Solids, 333, 2004, p. 44
[61] A. P. Webb and S. Veprek
Chem. Phys. Lett., 62, 1979, p. 173.
[62] A. Madan, S. R. Ovshinsky and E. Benn
Phil. Mag. B, 40, 1979, p. 259.
[63] R. Tsu, M. Izu, S. R. Ovshinsky and F. H. Pollak
Solid State Commun., 36, 1980, p. 817.
[64] S. Usui and M. Kikuchi
J. Non-Cryst. Solids, 34, 1979, p. 1.
[65] T. Hamasaki, H. Kurata, M. Hirose and Y. Osaka
Appl. Phys. Lett., 37, 1980, p. 1084.
[66] A. Matsuda et al.
Jpn. J. Appl. Phys., 20, 1980, p. 183.
[67] L.T. Canham
Appl. Phys. Lett., 57, N$^{o}$ 10, 1990, pp. 1046-1048.

[68]   Proc. Mater. Res. Society
       Microcrystalline and Nanocrystalline Semiconductors, Symp. F, Nov. 30-Dec. 3, 1998
       Session F1: Light Emission from Nano-crystalline Silicon
[69]   Y. Uchida, T. Ichimura, M. Ueno and H. Haruki,
       Jpn. J. Appl. Phys., 21, 1982, p. 586.
[70]   Y. Ikeda, T. Ito, Y. Li, M. Yamazaki, Y. Hasegawa and H. Shirai
       Jpn. J. Appl. Phys., 43, N$^o$ 9A, 2004, pp. 5960-5966.
[71]   S. Ghosh, A. Dasgupta and S. Ray
       J. Appl. Phys., 78, 1995, p. 3200.
[72]   H. Keppner, J. Meier, P. Torre, D. Fischer and A. Shah
       Appl. Phys. A, 69, 1999, p. 169 and references therein.
[73]   R. Biswas, Y. Ye and B.C. Pan
       Phys. Rev. Lett., 88, 2002, p. 205502.
[74]   T. Matsumoto, M. Kondo, S.V. Nair and Y. Masumoto
       J. Non-Cryst. Solids. 227-230, 1998, pp. 320-323.
[75]   J. P. Kleider, C. Longeaud, R. Bruggemann, and F. Houze
       Thin Solid Films, 383, 2001, pp. 57-60.
[76]   C.Y. Lin, Y.K. Fang, S.F. Chen, C.S. Lin, T.H. Chou, S.B. Hwang, J.S. Hwang, K.I. Lin
       J. of Electr. Mat., 34, N$^o$ 10, 2005, pp. 1123-1128.
[77]   A. Fejfar, N. Beck, H. Stuchlikova, N. Wyrsh, P. Torres, J. Meier, A. Shah and J. Kocka
       J. Non-Cryst. Solids, 227-230, 1998, p. 1006.
[78]   J. Perrin, Y. Takeda, N. Hirano, Y. Takeuchi and A. Matsuda
       Surf. Sci., 210, 1989, p. 114.
[79]   T. Ikeda, K. Sato, Y. Hayashi, Y. Wakayama, K. Adachi and H. Nishimura
       Solar Energy Materials & Solar Cells, 34, 1994, p. 379.
[80]   J. I. Pankov, P. J. Zanzucchi, C. W. Magee and G. Locpvsky
       Appl. Phys. Lett., 46, 1985, p. 421.
[81]   M. Kondo, Y. Nasuno, H. Mase, T. Wada and A. Matsuda
       J. Non-Cryst. Solids, 299-302, 2002, pp. 108–112.
[82]   R. Fluckiger, J. Meier, A. Shah, A. Catana, M. Brunel, H. V. Nguyen, R. W. Collins and R. Carius
       Proc. Mat. Res. Soc. Symp., 336, 1994, p. 511.
[83]   Y. Nasuno, M. Kondo and A. Matsuda
       Jpn. J. Appl. Phys., 40, 2001, p. 303.
[84]   Y. Hamakawa, H. Okamoto and Y. Nitta
       Appl. Phys. Lett., 35, 1979, p. 187.
[85]   S. Veprek, F. A. Sarott and M. Ruuckschloss
       J. Non-Cryst. Solids, 137-138, 1991, p. 733.
[86]   Y. Nasuno, M. Kondo and A. Matsuda
       Appl. Phys. Lett., 78, 2001, p. 2330.
[87]   J. Meier, H. Keppner, S. Dubail, U. Kroll, P. Torres, P. Pernet, Y. Ziegler, J. A. Selvan, J. Cuperus, D. Fischer and A. Shah
       Proc. Mater. Res. Soc. Symp., 507, 1999, p. 139.
[88]   A. Gallagher
       J. Appl. Phys., 60, 1986, p. 1369.

[89]   M. Kushner,
       J. Appl. Phys., 63, 1988, p. 2352.
[90]   S. Jones et al.
       Proc. Mater. Res. Soc., 297, 1993, p. 815.
[91]   L. Yang and L. F. Chen
       Proc. Mater. Res. Soc., 336, 1994, p. 669.
[92]   E. Srinivasan and G. Parsons
       Proc. Mater. Res. Soc., 467, 1997, p. 501.
[93]   G. Tao, M. Zeman and J. W. Metselaar
       Solar Energy Materials & Solar Cells, 34, 1994, p. 359.
[94]   F. G. Frieser
       J. Electrochem. Soc., 115, 1968, p. 401.
[95]   A. Barhdadi
       ICTP preprint, IC-2003, 102
[96]   A. Barhdadi
       Afrique Sciences, 1 (1), 2005, pp. 15-32.
[97]   O. P. Agnihorti and V. K. Rathi
       Semiconductor Materials and Devices, New Delhi, Narosa, 1998, p. 21
[98]   H. Ito, M. Hatanaka, K. Mizuguchi, K. Miyake and H. Abe
       Proc. Symp. Dry Process Institute of Electrical Engineers of Japan, Tokyo, 1982, p. 100
[99]   Y. Numasawa, K. Yanazaki, K. Hamano and F. Kobayashi
       Electrochem. Soc. Extended Abstracts, San Francisco, 1983, p. 662
[100]  V. K. Rathi, M. Gupta, R. Thangaraj, K. S. Chari and O. P. Agnihotri
       Thin Solid Films, 266, 1995, p. 219.
[101]  Kumar Vipan, K. S. Chari and O. P. Agnihotri
       Thin Solid Films, 232, 1995, p. 47.
[102]  V. K. Rathi, M. Gupta and O. P. Agnihotri
       Microelectron. J., 26, 1995, p. 563.
[103]  K. Nagamine, A. Yamada, M. Konagai, K. Takahashi
       Jpn. J. Appl. Phys. 26, 1987, p. L951.
[104]  F. Demichelis, C. F. Pirri and E. Tresso
       Philos. Mag. B 66, 1992, p. 135.
[105]  O. P. Agnihotri, S. C. Jain, J. Poortmans, J. Szlufcik, G. Beaucarne, J. Nijs, R. Mertens
       Semicond. Sci. Technol., 15 (7), 2000, pp. R29-R40.
[106]  P. K. Boyer, G. A. Roche, W. H. Ritchie and G. J. Collins
       Appl. Phys. Lett., 40, 1982, p. 716.
[107]  M. Tsuji, M. Sakumoto, N. Itooh, H. Obase and Y. Nishimura
       Appl. Surf. Sci., 51, 1991, p. 171.
[108]  R. Iyer and D. L. Lile
       J. Electrochem. Soc., 135, 1988, p. 691.
[109]  D. Bauerle, P. Irsiler, G. Leyenedecker, H. Noll and D. Wanger
       Appl. Phys. Lett., 40, 1982, p. 819.
[110]  K. Usami, Y. Mochizuki, T. Minagawa and A. Iida
       Jpn. J. Appl. Phys., 25, 1986, p. 1449.
[111]  T. L. Pollock, H. S. Sandhu, A. Jodhan, O. P. Strausz
       J. Am. Chem. Soc., 95, 1973, p. 1017.

[112] H. J. Emeleus, K. Stewart
Trans. Faraday Soc., 22, 1936, p. 1577.
[113] J. H. Clark, R. G. Anderson
Appl. Phys. Lett., 32, 1978, p. 46.
[114] K. Abe, T. Watahiki, A. Yamada, M. Konagai
Jpn. J. Appl. Phys., 38, 1999, p. 3622.
[115] T. Oshima, A. Yamada, M. Konagai
Jpn. J. Appl. Phys., 36, 1997, p. 6481.
[116] E. Kamaratos and F. W. Lampe
J. Phys. Chem., 24, 1970, p. 2267.
[117] E. R. Auastin and F. W. Lampe
J. Phys. Chem., 81, 1997, p. 1134.
[118] J. Perrin and T. Broekhuizen
J. Quant. Spectrom. Radiat. Transf., 38, 1987, p. 369.
[119] J. Perrin and T. Broekhuizen
Appl. Phys. Lett., 50, 1987, p. 433.
[120] P. John and J. H. Purnell
Faraday Trans., 69, 1973, p. 1455.
[121] J. M. Jasinski and J. O. Chu
J. Chem. Phys., 88, 1988, p. 1678.
[122] M. J. Kushner
J. Appl. Phys., 63, 1988, p. 2532.
[123] T. Fuyuki, B. Allain and J. Perrin
J. Appl. Phys., 68, 1990, p. 3322.
[124] A. Bouchoule, C. Laure, P. Ranson, D. Salah and D. Henry
Proc. 15th Symp. Plasma Processing, 1985, p. 399.
[125] J. Perrin and T. Broekhuizen
Proc. of Mater. Res. Soc. Symp., 75, 1987, p. 201.
[126] T. Toyoshima, K. Arai, A. Matsuda and K. Tanaka
Appl. Phys. Lett., 56, 1990, p. 1540.
[127] M. Tao
Thin Solid Films, 307, 1997, p. 71.
[128] R. P. Wayne
Comprehensive Chemical Kinetics, Elsevier, Amsterdam, Vol. 2, 1969, Chap. 3.
Editors: C. H. Bamford and C. F. H. Tipper
[129] P. C. Riedi,
Thermal Physics: An Introduction to thermodynamics, Statistical Mechanics and Kinetic Theory, 2$^{nd}$ Edition, Oxford University, Oxford, 1988, Chap. 8.
[130] K. Kamisako, T. Imai and Y. Tarui,
Jpn. J. Appl. Phys., 27, 1988, p. 1092.
[131] K. Suzuki, K. Kuroiwa, K. Kamisako and Y. Tarui,
Appl. Phys. A, 50, 1990, p. 277.